\documentclass{PoS}

\title{Radio-detection of extensive air showers at the Pierre Auger Observatory -- Results and enhancements}

\ShortTitle{Radio-detection of extensive air showers at the Pierre Auger Observatory}

\author{\speaker{Karim Louedec}$^1$, for the Pierre Auger Collaboration$^2$
	\thanks{http://www.auger.org/archive/authors\_2013\_06.html}\\
        $^1$ Laboratoire de Physique Subatomique et de Cosmologie (LPSC), UJF-INPG, CNRS/IN2P3, Grenoble, France.\\
        $^2$ Observatorio Pierre Auger, Av. San Mart\'in Norte 304, 5613 Malarg\"ue, Argentina.\\
        E-mail: \email{karim.louedec@lpsc.in2p3.fr}}

\abstract{The Pierre Auger Collaboration is exploring the potential of radio-detection techniques to measure the extensive air showers. The main advantage of these setups is the possibility to cover a large area with no atmospheric attenuation and having 100\% duty cycle. Radio emission in the MHz range is recorded by the Auger Engineering Radio Array (AERA), presently consisting of 124 stations distributed over an area of approximately 6~km$^2$. This enhancement is focused on physics of cosmic rays with an energy greater than 10$^{17}$~eV. In addition, novel detection techniques based on the GHz emission from extensive air showers are being tested at the Pierre Auger Observatory. Three different setups are currently installed and are collecting data: MIDAS (Microwave Detection of Air Showers) and AMBER (Air-shower Microwave Bremsstrahlung Experimental Radiometer) are prototypes of an imaging parabolic dish detector, while EASIER (Extensive Air Shower Identification using Electron Radiometer) records the radio emission by antenna horns located on more than 60 surface detector units. The status of these different activities and the new results in MHz and GHz bands will be reported.}

\FullConference{The European Physical Society Conference on High Energy Physics \\
		 18-24 July, 2013\\
		 Stockholm, Sweden}

\begin{document}

\section{Introduction}
\label{sec:intro}
The Pierre Auger Observatory is the largest operating cosmic ray observatory ever built~\cite{OPA}. It is designed to measure the flux, arrival directions and mass composition of cosmic rays from above $3\times 10^{17}~$eV to the very highest energies. The observatory is composed of 1660 water Cherenkov detectors covering 3000~km$^2$, and 27 fluorescence telescopes surrounding the array on four sites. Both of these techniques have some limitations: whereas the interpretation of the Cherenkov detector data is very dependent on hadronic interaction models which are not yet tested in laboratories at the highest energies, the optical telescopes are operated only during dark nights (duty cycle close to $13\%$). It is in this context that the Pierre Auger Collaboration is pursuing R\&D projects to enhance the detection capabilities and to test new techniques for the next generation of ground-based detectors. Currently, the most promising technique is probably the radio-detection with a duty cycle reaching almost $100\%$ and no atmospheric attenuation. This technique allows one to record the whole development of the extensive air shower with a quasi-calorimetric measurement of the shower energy. Two different frequency domains are probed at the Pierre Auger Observatory: the MHz band and, more recently, the GHz band. In the following, we give a status on both activities and their next challenges.

\section{The Auger Engineering Radio Array (AERA) -- MHz frequencies}
\label{sec:mhz}
The Auger Engineering Radio Array (AERA) records radio signals produced by extensive air showers in the MHz frequency domain, from 30 to 80~MHz~\cite{AugerRadio_FAL}. After a first phase, started in April 2011 with 24 stations distributed over an area of 0.5~km$^2$, the second phase is operating since April 2013 with 100 additional stations covering an area of around 6~km$^2$. It is planned to deploy an additional 36 stations in 2014 to cover, finally, approximately 20~km$^2$. Although AERA can be operated in a self-trigger mode, a portion of the radio stations can get an external trigger from the water Cherenkov detectors or from scintillators integrated directly into the stations. In this MHz frequency range, there is a polarised coherent emission resulting from two relevant mechanisms: {\it the geomagnetic effect} corresponding to the transversal separation of charged particles due to the geomagnetic field (linear polarisation), and the {\it Askaryan effect} related to the variation of charge excess along the longitudinal development of the air shower (radial polarisation). Both of them have been confirmed by the Pierre Auger Collaboration using events recorded during the first phase of AERA~\cite{Melissas,HuegePolar}. Using multi-hybrid events and Monte Carlo simulation codes dedicated to radio emission, we are improving our understanding of radio emission mechanisms. Current studies are focused on evaluating the precision of AERA for estimating the arrival direction, energy and composition of air showers in the energy range from $10^{17}$ to $10^{19}~$eV.

\section{Microwave radiation (AMBER, EASIER, MIDAS) -- GHz frequencies}
\label{sec:ghz}
The observation of a microwave continuum emission from air shower plasmas has raised the idea to measure extensive air showers in the GHz frequency range~\cite{Gorham}. This recorded signal was interpreted as molecular bremsstrahlung radiation (MBR), {\it i.e.}\ an emission produced by low-energy electrons (coming from ionisation of air molecules by primary electrons of the air shower) scattering into the electromagnetic field of neutral atmospheric molecules. Radio emission produced by this mechanism presents interesting features: it is isotropic and unpolarised, with a very low natural background. Three prototypes, AMBER, EASIER and MIDAS are now operating at the Pierre Auger Observatory to measure microwave radiation coming from extensive air showers~\cite{RGaior}. The main goal is to characterise this microwave emission and to test the feasibility of such a technique to detect ultra-high energy cosmic rays. AMBER and MIDAS are imaging telescopes composed of horn antennas placed at the focus of a parabolic dish. This design allows one to record the shower longitudinal profile by exploiting the isotropic nature of the MBR. EASIER instruments a set of 61 water Cherenkov tanks with a microwave receiver observing the shower with a wide field-of-view (around $100^\circ$) pointing to the zenith. In this set-up, the very small effective area of the antennas is compensated by time compression of the signal and closer distances to the shower. All of them record signals in the C-band ($3.4-4.2~$GHz) and AMBER has additional horn antennas in the Ku-band ($10.9-14.5~$GHz). The MIDAS detector, during a first phase of data-taking at the University of Chicago, sets a limit on air shower microwave emission and excludes a quadratic scaling with the air shower energy (for an isotropic emission)~\cite{MIDAS}. Up to now, only the EASIER project recorded microwave signals in coincidence with an extensive air shower, especially with antennas close to the shower core and with an E-W polarisation~\cite{RGaior}. Several analyses to understand the origin of these signals are on-going.

\section{Conclusion and outlook}
\label{sec:conclusion}
The Pierre Auger Observatory is a great facility for studying alternative air shower detection techniques. Whereas the collaboration is now working on the MHz radio emission to estimate properties of cosmic rays, we are still in an exploratory phase to identify the main mechanism(s) producing signals in the microwave band.

\end{document}